\documentstyle{l-aa}
\input epsf

\def\rh{r_h}
\def\mv{M_V}
\def\mbol{M_{\rm bol}}
\def\814{I_{814}}
\def\606{R_{606}}
\def\555{V_{555}}
\def\mh{[M/H]}
\def\feh{[Fe/H]}
\def\msol{M_\odot}
\def\lsol{L_\odot}

\def\te{T_{\rm eff}}

\def\wig#1{\mathrel{\hbox{\hbox to 0pt{%
          \lower.5ex\hbox{$\sim$}\hss}\raise.4ex\hbox{$#1$}}}}
\newcommand{\msvol}{\mbox{M}_\odot\, \mbox{pc}^{-3}}

\renewcommand{\thetable}{\Roman{table}}

\begin{document}

\def\aj{AJ}                  
\def\araa{ARA\&A}             
\def\apj{ApJ}                 
\def\apjl{ApJ}                
\def\apjs{ApJS}               
\def\ao{Appl.Optics}          
\def\apss{Ap\&SS}             
\def\aap{A\&A}                
\def\aapr{A\&A~Rev.}          
\def\aaps{A\&AS}             
\def\baas{BAAS}               
\def\jrasc{JRASC}             
\def\memras{MmRAS}            
\def\mnras{MNRAS}             
\def\pra{Phys.Rev.A}          
\def\prb{Phys.Rev.B}          
\def\prc{Phys.Rev.C}          
\def\prd{Phys.Rev.D}          
\def\prl{Phys.Rev.Lett}       
\def\pasp{PASP}               
\def\pasj{PASJ}               
\def\qjras{QJRAS}             
\def\skytel{S\&T}             
\def\solphys{Solar~Phys.}     
\def\sovast{Soviet~Ast.}      
\def\ssr{Space~Sci.Rev.}      
\def\zap{ZAp}                 

\thesaurus{}

\title{Determination of the globular cluster and halo stellar mass functions
and stellar and brown dwarf densities}

\author{{\sc Gilles Chabrier and Dominique M\'era}}

\institute{
C.R.A.L. (UMR 7455 CNRS),\\
Ecole Normale Sup\'erieure, 69364 Lyon Cedex 07, France\\
(chabrier, dmera @ens-lyon.fr)}

\date{Received date : 10 January, 1997 ; accepted date : 28 April, 1997}

\maketitle

\markboth{G. Chabrier and D. M\'era : Globular cluster and halo stellar mass functions} {}

\begin{abstract}

\bigskip
We use recent low-mass star models, which reproduce accurately
the observed sequences of various globular clusters, to convert the observed luminosity functions into bolometric {\it luminosity functions}. These latter are shown to exhibit a similar behaviour with a rising slope up to $\mbol \sim 9$,
i.e. $m\sim 0.2-0.3\,\msol$, and a decreasing behaviour beyond this limit. We then derive {\it mass functions} for globular star clusters down to the bottom of the main sequence. These mass functions are well described
by a slowly {\it rising} power-law $dN/dm\propto m^{-\alpha}$, with $0.5\wig < \alpha \wig < 1.5$, down to $\sim 0.1\,\msol$, independently of the metallicity, suggesting a rather
universal behaviour of the cluster initial mass functions. The effects of tidal stripping
and mass segregation are illustrated by the overabundance of very low-mass
stars in the outer parts of $NGC6397$ and their depletion in the central parts. This analysis confirms that the mass function determined near the
half-mass radius has been weakly affected by external and internal dynamical effects and reflects relatively closely the {\it initial} mass function.
We predict luminosity functions in the NICMOS filters in the stellar and in the
brown dwarf domains for different mass functions and metallicities.

We apply these calculations to the determination of the mass function in the
Galactic halo, including the DeVaucouleurs spheroid and the $1/r^2$ dark halo.
We derive the slope and the normalization of the spheroid mass function which is well described by the afore-mentioned power-law function with $\alpha \sim 1.7\pm 0.2$ down to
0.1 $\msol$, although
a slowly decreasing mass function below $\sim 0.15\,\msol$ can not be excluded
with the data presently available. Comparison with the Hubble Deep Field
star counts is consistent with such a mass function and excludes a significant
stellar population in the dark halo. This shows that essentially all the high-velocity
subdwarfs observed in the solar neighborhood belong to the Galactic spheroid. 

Consistent analysis with recent microlensing experiments towards the LMC shows
that the spheroid and the dark-halo {\it stellar+brown dwarf} populations
represent at most $\sim$ 1\% of the Galactic dark matter density. This clearly excludes
brown dwarfs and low-mass stars as significant dark matter candidates.
\bigskip

\end{abstract}

\section{Introduction}

The derivation of reliable stellar mass-functions (MF) down to the hydrogen
burning limit is an essential task for the understanding of a wide variety
of major astrophysical problems including star formation theory, stellar
evolution, dynamical evolution of globular clusters, galactic formation and evolution. Not mentioning the long-standing problem of the galactic missing mass. From the stellar point of
view, the knowledge of the low-mass end of the MF provides essential information
about the processus of star formation and the physics of the interiors
and atmospheres of cool and dense objects. From the galactic point of view,
the determination of the stellar MF down to the hydrogen burning limit
is a key issue
to derive the total amount of matter under the form of visible,
stellar objects, which determines the mass-to-light ($m/L$) ratio, and to infer
the contribution of substellar objects, essentially brown dwarfs,
to the galactic mass budget. The intrinsic faintness of very-low-mass stars (VLMS), and the small density of halo objects w.r.t.
the disk population ($\sim 1/500$, see e.g. Bahcall, 1986; Tyson, 1988) renders the determination of the luminosity function (LF) of
halo stars down to the hydrogen burning limit a tremendously difficult task.
The determination of globular cluster (GC) luminosity functions and mass
functions is a first, easier step towards the determination of the Population II star initial
mass function. Moreover, the determination of the present day
MF (PDMF) in GCs contains potential information about the dynamical history, evaporation
and tidal stripping of the clusters, and thus about the Galactic halo history.
At last, given the large range of metallicities covered by all observed
GCs, it is possible to examine the dependence of the MF on
metallicity, and to determine whether it is a universal property or
if it is strongly affected by the metal content of the initial cloud.

Recent deep photometry observations of several GCs with the Hubble Space Telescope
(HST) yield an accurate determination of the bottom of the MS of these objects.
These results extend previous
ground-based observations to about 4 magnitudes fainter limits and allow an accurate determination
of the cluster LFs down to the bottom of the MS, providing stringent constraints
for theoretical models of metal-poor VLMS.
The derivation
of accurate MFs from these observed LFs is a more delicate task. The MF is
very sensitive to the mass-luminosity relation and thus the derivation of
reliable MFs requires the calculations of accurate mass-luminosity relations.
Direct, observational mass determinations of VLMS are available for solar metallicity objects only
(Popper 1980; Henry \& McCarthy 1993). No direct mass-determination
has been obtained so far for metal-depleted low-mass stars. However the accuracy of
the stellar models
can be examined by confronting the theoretical color-magnitude diagrams (CMD) to the observed
ones. Such an extensive study has been conducted in a previous paper
(Baraffe, Chabrier, Allard \& Hauschildt 1997; BCAH) where these authors have derived
an extended set of low-mass star evolutionary  models ($0.07\le m\le 0.8\,\msol$) over a
large range of metal-depleted abundances characteristic of old disk, globular cluster
and halo populations, namely $-2.0\le \mh \le -0.5$. The accuracy of these
models has been demonstrated by the {\it remarkable agreement} ($< 0.1$ mag) between the theoretical
and the observed CMDs over the entire sequences,
for {\it all} GCs, over the entire afore-mentioned
metallicity range. The well known wavy behaviour of the stellar main sequence, which reflects genuine intrinsic
properties of stellar matter, is reproduced qualitatively and {\it quantitatively},
for the correct magnitudes and colors (see BCAH). As stressed by these authors, these models are not
hampered by {\it any} adjustable parameter (e.g. the mixing length) and thus the agreement
between theory and observation reflects the accuracy of the physics entering the models (equation of state and {\it non-grey}
atmosphere models). This assesses the reliability of these models to
describe the mechanical and thermal properties of metal-poor low-mass
stars down to the brown dwarf limit, and provides for the first time accurate
relations between the mass, the age and observable quantities, magnitudes and
colors, for different metallicities.
For solar-like metallicity, the accuracy of the mass-magnitude relation has been
assessed by the agreement with the observed mass-spectral class (Kirkpatrick \&
Mc Carthy, 1994; Kirkpatrick et al., 1995) and mass-$M_V$ and mass-$M_K$ (Henry \& McCarthy, 1993)
relationships (Baraffe \& Chabrier, 1996; Chabrier, Baraffe \& Plez, 1996; Chabrier \& Baraffe, 1997b).

In this paper, we use the mass-luminosity relationships obtained
with the afore-mentioned metal-depleted models to derive the mass-functions characteristic of several
GCs observed with the HST and of halo field stars. The observations are summarized in \S 2 and converted into {\it bolometric} LFs.
In \S 3, we derive
the MFs for the different clusters and we examine the dependence of these MFs
upon metallicity, location in the cluster and Galactic position of the cluster.
Section 4 is
devoted to the derivation of the {\it halo} MF and \S 5 to the conclusion.

\section{Luminosity Functions}

Until recently, derivations of stellar LFs in GCs down to faint absolute magnitudes have been based on observations
made from the ground, thus suffering from crowding effects and uncertain completeness corrections (see e.g. Richer et al. 1991). Data from the
HST have now become available for several GCs down to $\sim 4$ 
magnitudes fainter limits ($M_V\sim 14$) without the need for large completeness corrections, spaning
a large range of metallicity from solar to 1\% solar.
The MS is sharp and well defined down to almost the limit
of the observations, which corresponds to $m\sim 0.1\,\msol$ (BCAH), and the
LFs are essentially complete (within $\sim 80\%$)
down to this limit. These observations are
presented and discussed in \S 2 of BCAH, with
the different parameters, distance moduli and reddening corrections, used for these clusters.

\begin{table*}
\caption[]{Metallicity, distance moduli (corrected from extinctions; see caption of Fig. 1) and Galactic location of the globular clusters.
$[Fe/H$] is the observed metallicity, $[M/H]$ is the metallicity used in the models, which takes into
account the enrichment in $\alpha$-elements (see BCAH for details).
Z$_{GP}$, R$_{GC}$ and r/r$_h$ denote the height above or below the Galactic disk, the distance from the Galactic center and the location of the observations w.r.t. the half-mass radius, respectively.}
\begin{tabular}{llllllll}
\hline\noalign{\smallskip}
$cluster$ & $[Fe/H]$ & $[M/H]$   
& $(m-M)_I$ & $(m-M)_V$ & Z$_{GP}$ (kpc) & R$_{GC}$ (kpc) & $\sim$r/r$_h$ \\ \noalign{\smallskip}
\hline\noalign{\smallskip}
 47Tuc$^{1,2}$ (NGC104)     & -0.7    & -0.5 & 13.35/13.46 & & 3.3 & 8.0 & 0.8  \\
 NGC6752$^3$      & -1.5    & -1.0   & 13.11   &      & 1.8  & 5.9  & 1.3   \\
 M3$^{3}$ (NGC5272)        & -1.3    & -1.0 & 15.07 &   & 10.8& 12.5  & 3.7 \\
 $\omega$ Cen$^4$ (NGC5139) & -1.6  & -1.3 & 13.92 & & 1.3 & 7.0 & 3.2-4.6 \\
 NGC6397$^{5,6}$  & -1.9    & -1.5 & 12.05/12.20 & 12.47 & 0.5 & 6.0 & 3  \\
 NGC6656$^{8}$  & -1.75    & -1.5 &  &  & 0.5 & 6.0 & 3  \\
 M30$^{5}$ (NGC7099)       & -2.1    & -1.8 & 14.48 & 14.65 & 5.25& 7.0 & 3   \\
 M15$^{5,7}$ (NGC7078)      & -2.2    & -2.0 & 15.25 &15.4   & 4.5 & 10.5& 3 \\
 M92$^{5}$ (NGC6341)       & -2.2    & -2.0 & 14.58 & 14.6  & 4.4 & 9.8 & 3 \\ \noalign{\smallskip}
\hline
\end{tabular}
\bigskip

$^1$ De Marchi \& Paresce 1995a

$^2$ Santiago et al., 1996

$^3$ Fusi Pecci et al., 1996

$^4$ Elson et al. 1995

$^5$ Cool et al. 1996

$^6$ Paresce et al. 1995

$^7$ De Marchi \& Paresce 1995b

$^8$ De Marchi \& Paresce 1997

\end{table*}

Although there is a general agreement between the ground-based and
the HST LFs at bright and intermediate magnitudes, they differ substantially at and
beyond the magnitude limit of ground-based observations. Although the
ground-based LFs were generally rising down to the limit of the observations
($M_I\sim 9$),
most of the LFs derived with the HST show a clear decline beyond this limit.

A comparison between the LFs of some of the afore-mentioned clusters, namely
$47Tuc$, $NGC6397$ and $\omega Cen$, has been done by Cool et al.
(1996) and Santiago  et al. (1996). In spite of the differences
between these clusters arising from their age, metallicity, galactic position,
and thus dynamical evolution, the LFs are found to be very similar, down to the limit of the observations,
except for the faint end of $\omega Cen$.

The clusters have been observed in different passbands characteristic of the HST,
namely $F814\sim I_C$, $F555\sim V_J$ and F$606$ where $I_C$ and $V_J$ denote the standard
Johnson-Cousins (JC) system. HST colors were transformed into the JC system, using the
Holtzman et al. (1995) synthetic transformations. As shown in BCAH, agreement between the theoretical and the observed
CMDs
is remarkable in {\it all} filters, HST and JC, assessing both the accuracy of the models
and of the Holtzman photometric transformations (see BCAH for details).
In Fig. 1, we show the {\it bolometric} LFs, corrected for incompleteness, for the afore-mentioned
clusters, using the published LFs and the bolometric corrections of Chabrier \& Baraffe (1997a) and
BCAH.
We have
used cubic spline interpolations to interpolate the relations M$_\lambda(\mbol)$
and their
derivatives dM$_\lambda$/d$\mbol$ (global polynomial fits are {\it not} reliable). We verified that LFs of the
same cluster observed in {\it different passbands} yield the {\it same} bolometric LF. This adds
credibility to the present bolometric corrections and stellar models.

\begin{figure}
\epsfxsize=85mm
\epsfysize=100mm
\epsfbox{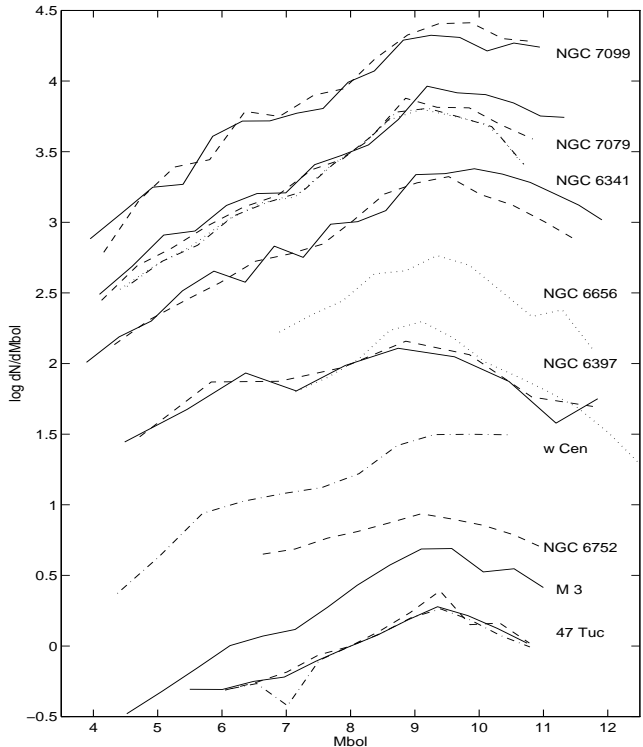}
\caption[]{
Bolometric luminosity functions of different globular clusters, obtained from the observed
LFs in various passbands, including 
$F814$, $F555$ and $F606$, using the
bolometric corrections of Baraffe et al. (1997; BCAH) and Chabrier \& Baraffe (1997a). The clusters are displayed by {\it increasing metallicity} (see Table I) from top to bottom, namely : $NGC7099$, $NGC7078$, $NGC6341$ from Cool et al. (1996) observations in the $V$ (solid line) and $I_{814}$ (dashed line) bands and from DeMarchi \& Paresce (1995b) in $I_{814}$ (dotted line); $NGC6656$ from DeMarchi \& Paresce (1997) in $I_{814}$;
$NGC6397$ from Cool et al. (1996) and from Paresce et al., (1995b), same caption as above; $wCen$ from Elson et al. (1995) (dot-dash line); $NGC6752$ (dash) and $M3$ (solid) from Fusi Pecci et al. (1996);
$47Tuc$ from De Marchi \& Paresce (1995a) (solid line) and from
Santiago et al. (1996) (dot-dash line).
The distance moduli are given in Table I, corrected from
extinctions with the corrections given in Table I of BCAH or
quoted by the respective observers. All observations
are normalized arbitrarily at $\mbol =8$ for each cluster. The error bars (see the afore-mentioned references) are not shown for a sake of clarity.}

\end{figure}

We believe these results to illustrate
consistent
comparisons between the LFs of various clusters, over a large range of metallicities, from near solar to
$\mh=-2.0$.
The LFs have been normalized arbitrarily to $\mbol=8$, i.e. $m\sim
0.4-0.5\,\msol$ over the
present range of metallicity.
For all the considered GCs, both the degree of completeness and of photometric accuracy are maximum
in this region and stellar evolution effects do not affect this mass range, even for the older clusters. 
In Table I, we give the galactocentric position $R_{GC}$, the height above or below the Galactic plane $Z_{GP}$ and the metallicity for the various clusters. These values have been averaged over the slightly
different values found in the literature. We also give the location of the observations w.r.t. the half-mass radius $r_h$.
Some indetermination ($\sim 0.2 $ mag) remains for the distance modulus of various clusters, as quoted by the different observers.
We have carried out calculations with the different values and found that the
resulting uncertainty in the MFs remains within the observational error bars.
The extinction corrections are the ones quoted in Table I of BCAH and in the
different references corresponding to the observations.

As seen in Fig. 1, all LFs exhibit a similar qualitative behaviour
in the region
$\mbol \wig < 9$, i.e.
$M_{814}\approx M_I\wig < 8.5$, $M_{555}\approx M_V\wig < 10$, with a rising
slope $\Delta \log (dN/d\mbol)/ \Delta \mbol \sim 0.25$, i.e.
$dN/d\log (L/\lsol) \propto -(L/\lsol)^{-0.6}$.
Below this limit, some
differences appear between the various clusters. Whereas some LFs, e.g. for
$\omega Cen$, exhibit a flattish behaviour, most of them drop more or less
rapidly at the faint end.
As will be shown in the next section, these different behaviours, and the respective magnitudes
corresponding to the {\it peak} in the LFs of the clusters, reflect the
dependence of the mass-magnitude relationship on metallicity, from a rather
universal mass-function.
The effect of metallicity in the mass-magnitude relationship
translates into a slightly different location for the peak of the luminosity
function, corresponding to $m\sim 0.2\,\msol$, lower metallicities yielding
brighter magnitudes for the same mass, an intrinsic
property of stellar matter (see BCAH).
The case of $NGC6397$ deserves particular attention, for 
there is a substantial {\it qualitative} difference in
the faint end of the LF for this cluster between the observations of Paresce et al. (1995b) and Cool et al. (1996).
Whereas Paresce et al.
get a strongly decreasing LF, Cool
et al. argue for a milder decrease, or even for a flattish behaviour at the faint end of the LF. Since both observations were completed in a similar region of the cluster ($r/\rh\sim 3$), the difference can not stem from different mass segregation effects. A natural explanation would be incompleteness at faint magnitudes in the Paresce et al. 
sample but this hypothesis seems to be excluded by these authors.
A third explanation would be the bias in the observed LF due to bining at faint magnitudes.
This point will be discussed in detail in \S 4.
In any case, this uncertainty
prevents an accurate determination of the mass function (MF) at the bottom of the MS for $NGC6397$,
and stresses the need for further observations of this cluster.

These various LFs were obtained at similar location in the clusters,
between $\sim 1$ and 3 times the half-mass radius (see Table I), assuming that the present half-{\it light}
radius is
close to the half-mass radius, a reasonable assumption (Djorgovski, 1993).
In any case the region where the LFs have been determined is
well inside the tidal radius. As discussed below, this is the best location
to obtain a PDMF as similar as possible to the {\it initial} MF (IMF), although
dynamical evolution may have shaped it to some extent.

In summary, the different LFs seem to be marginally similar down
to $\mbol \sim 10$, in spite of the very different metallicities. Whether this
apparent non-dependence of the LF upon metallicity will translate into a
similar uncorrelation in the {\it mass-function} depends on the
degree of correlation of the {\it mass-luminosity} relationship with metallicity,
as examined below.

\section{Globular cluster mass-functions}

As already mentioned, the agreement between the observed GC main sequences and the
theoretical BCAH ones over the {\it entire} sequences
is excellent, less than 0.1 mag, for all
clusters.
In particular the models reproduce accurately the two changes in the slope
of the color-magnitude relation, near $\sim 0.4-0.5\,\msol$ and below $\sim 0.2\,\msol$,
which reflect
intrinsic physical properties of the atmosphere and the interior of VLMS
(see Chabrier \& Baraffe, 1997a). 
These comparisons
asses the validity of the
calibration of low-mass stars from the observed colors and magnitudes, as given in
BCAH and in Chabrier \& Baraffe (1997a).

As demonstrated in the afore-mentioned papers,
relationships based on {\it grey} atmosphere models and $\te(\tau)$ relationships
yield inaccurate, or at best unreliable, $m-L$ and $m-\te$ relationships over the entire low-mass star
domain, and most importantly near the hydrogen burning limit. This translates directly into unreliable
MFs,
thus biaising any attempt to derive robust conclusions.
We stress also the importance of using an accurate representation of the $m-L$ relation
when deriving mass functions. Given the severe variations of the slope of
this relation along the LMS range, a global polynomial fit, in particular, is a very poor and misleading representation.
A polynomial fit does {\it not} reproduce all the points of the $m-L$ relation
and can lead to severely erroneous MFs, in particular near the bottom of the MS.

\begin{figure}
\epsfxsize=88mm
\epsfbox{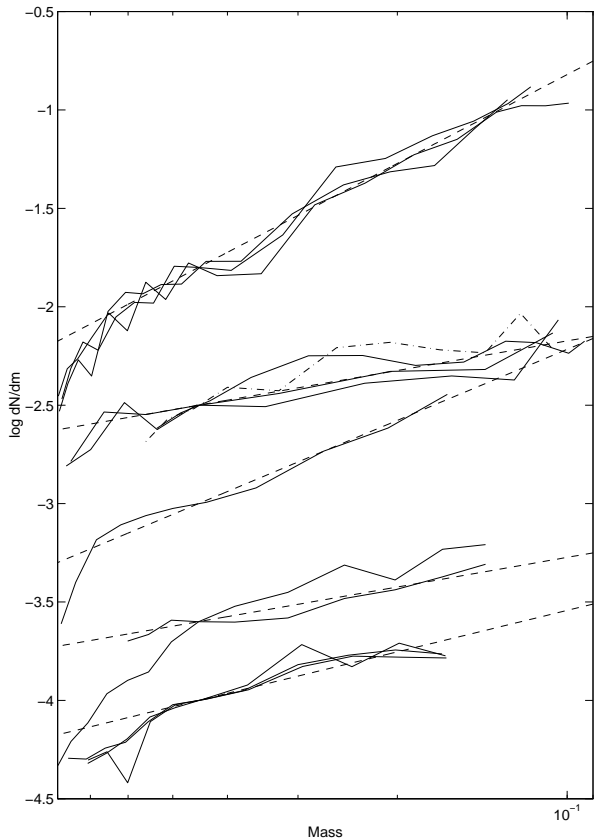}  
\caption[]{Mass-functions (in solar mass) for the different clusters with increasing metallicity from top
to bottom. $\alpha$ indicates the slope of the power-law $dN/dm\propto m^{-\alpha}$, as given below in brackets and shown by the dotted line for each cluster. Top to bottom :  $NGC6341$, $NGC7079$ and $NGC7099$ ($\mh \sim -2.0$; $\alpha \sim 1.5$); $NGC6656$ (dash-dot) and $NGC6397$ (solid) ($\mh =-1.5$; $\alpha \sim 0.5$); $wCen$ ($\mh\sim$ -1.3; $\alpha \sim 1.0$);
$NGC6752$ and $M3$ ($\mh \sim -1.0$; $\alpha \sim 0.5$);
$47Tuc$ ($\mh\sim -0.5$; $\alpha \sim 0.7$).
Several lines for a same cluster indicate different observations (see Fig. 1).
}
\end{figure}

The LFs discussed in \S 2 have been converted 
into MFs, for the appropriate
metallicities, and are displayed on Fig. 2.
As shown in Fig. 2, {\it all} MFs are rising monotonically down to the limit
of the observations, i.e. $\sim 0.1-0.15\,\msol$, close to the
hydrogen burning limit, except possibly for $47Tuc$ where the MF flattens below
$\sim 0.2\,\msol$. This shows convincingly that the decrease of the luminosity
function near the faint end does not reflect a lack of stars but instead the
{\it severe drop in the mass-luminosity relationship near the bottom of the MS} and thus the scarcity of stars over large magnitude intervals for $m\wig < 0.2\,\msol$
(see e.g. D'Antona \& Mazzitelli, 1996; Chabrier \& Baraffe, 1997a).
We stress again that LFs observed in
different filters, for a same cluster, yield similar MFs.
The MFs shown in Fig. 2 for the different clusters are displayed by increasing metallicity
from top to bottom, from $\feh \sim -2.2$ to $\feh\sim -0.7$.
We recall that, given the enrichment of halo and globular star clusters in
$\alpha$-elements, an observed $\feh$ for metal-poor stars corresponds to a solar-mix scaled metallicity
$\mh \sim \feh + 0.35$ (Ryan \& Norris, 1991; see BCAH for details).
A first interesting result is that,
except for the MF of $NGC6397$ derived from the Paresce et al. (1995b) LF, {\it all} MFs are
fairly well represented by simple power laws $dN/dm \propto m^{- \alpha}$ with $\alpha >0$. The second striking result
is the weak variation of the slope of the MF with metallicity, with $\alpha \sim 0.5$ to 1.5.
This dependence on metallicity is pondered to some extent by the location of the clusters
in the Galaxy.
This influence, however, remains weak as long as the clusters are located far enough
($\wig > 1$ kpc) from both the disk and the Galactic center, as shown for example for $NGC6752$ or
$wCen$, located about 1 kpc above the Galactic plane, which have a slope similar to or steeper than $M3$, of comparable metallicity, located $\sim 10$ kpc above the plane. In the same vein, $M30$ is found to have a slighly steeper MF than $NGC6341$, for the same metallicity,
although it lies closer to the Galactic center. This shows that there is no direct link between the MF near $r_h$ and the location in the Galaxy and that
some {\it internal} dynamical process, namely mass segregation, may have slightly modified
the IMF of the clusters near the half-mass radius. Although the {\it last bin} of $NGC6656$ seems to indicate a decrease at the faint end of the MF, this feature is likely to stem from incompleteness due to crowding near the bottom of the MS (DeMarchi \& Paresce, 1997). Interestingly enough, however, both the MF of
$NGC6656$ and $NGC6397$, as observed by DeMarchi \& Paresce, seem to indicate a
flattening below $\sim 0.2-0.3\,\msol$ (although the error bars in NGC6656 are consistent with a slowly rising MF), 
as suggested by these authors. Even though both clusters have similar Galactic positions (see Table I), they have quite different orbits and thus have experienced different dynamical histories
(DeMarchi \& Paresce, 1997). This prompted these authors to suggest that this
behaviour is an intrinsic properties of GC's. We note, however, that this result
differs from the ones obtained for all the other clusters and more importantly
from observations of the same cluster ($NGC6397$) by Cool et al. (1996) (see below), as mentioned previously.
Solving this issue requires further observations of these clusters near
the bottom of the MS.  
The case of $47Tuc$ is also of particular interest, for the MF of this relatively metal-rich cluster seems to flatten off below $\sim 0.2\,\msol$, although
the lack of observations beyond this limit prevents robust conclusions. The detailed study of the MF of metal-rich clusters and field stars is presently under work (Chabrier, M\'era \& Baraffe, 1997).

The similitude between the MFs of the different clusters, all determined at
radial distances near the half mass radius, seems to confirm that the
stellar population in this region is not very sensitive at least to
external dynamical processes.
In particular, given its large distance from the Galactic plane
and from the Galactic center,
$M15$ is believed to have
an almost uncontaminated MF, and is thus of particular importance for the
understanding of the dynamical evolution of GCs. It is instructive to note
that the MF of $M15$ (and $M3$) is
very similar to the one of clusters located closer to the Galactic plane,
although these clusters are expected to have
experienced quite different dynamical histories,
in particular very different tidal stripping and disk/bulge shocking,
resulting in a substantially different fraction of VLMS at the periphery.

\begin{figure}
\epsfxsize=88mm
\epsfbox{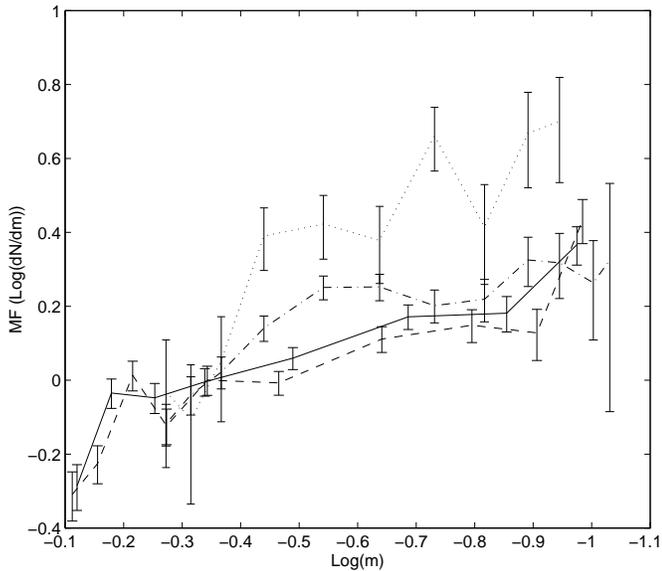}
\caption[]{ Mass function for $NGC6397$ from different observed LFs :
solid line : $I_{814}$ (Cool et al., 1996); dashed line : $V_{555}$ (Cool et al., 1996); dot-dashed line : $I_{814}$ (Paresce et al., 1995). Also shown is the MF obtained from the LF observed by Mould et al. (1996) at $r\sim 6 \times r_h$
(dotted line)
to illustrate mass segregation effects. The MFs are normalized at 0.5 $\msol$.
}
\end{figure}

As mentioned in the previous section, the indetermination near the faint end of the LF for
$NGC6397$ translates into a difference in the MF below $\sim 0.3\,\msol$. This is illustrated in Figure 3.
The {\it two} LFs of Cool et al. (1996) (in the $F555$ and $F814$ filters) yield a monotonic, rising MF, whereas the LF
of Paresce et al. (1995) yields a flattish MF below 0.3 $\msol$, although with large error bars near the faint end.
The difference remains when using any of the two distance moduli quoted for
this cluster and stems
from the very difference at the faint end of the two LFs. For this reason, we will leave this cluster aside for the
rest of the discussion until more observations can nail down this issue.
Note that
the low galactic position of $NGC6397$, $Z\approx 0.5$ kpc, its proximity to the Galactic center, and its likeky
collapsed state suggest that this cluster has experienced severe disk and bulge shocking,
resulting in an accumulation of low-mass stars at the periphery and a
substantial depletion even near $r_h$.
Note also that the relatively short half-mass
relaxation time
of NGC6397 ($t_h\sim 3\times 10^8$ yr; Djorgovski et al., 1993) may have led  already to substantial evaporation in this region.
This might provide a
simple explanation for the MF inferred from Paresce et al. data
and to explain the fact that $NGC6397$ has a shallower MF, and thus a smaller amount of LMS than clusters of similar metallicity, e.g. $M30$. On the opposite, if the MF keeps rising near the bottom of the MS, as inferred from the
Cool et al. LF, the similarity of the MF with the ones of
clusters located far from
the Galactic plane shows that even strong tidal stripping and disk
shocking does not {\it strongly} affect the inner regions of the cluster, near $\rh$.

An interesting information about tidal stripping and mass segregation in $NGC6397$ arises from the observations of Mould et al. (1996)
who obtained the LF of the cluster with the HST at a larger radial distance than the afore-mentioned observations,
namely at 10 arcmin from the center, i.e. $r/r_h\sim 6$. The MF obtained from Mould et al. LF
is shown in Fig. 3 (dotted line). The steeper slope ($\alpha \sim 1.3$ instead of $\alpha \sim 0.5$) and thus the larger number of very low-mass stars ($\wig < 0.5\,\msol$)
w.r.t. to the previous determinations closer to $r_h$ illustrates the onset of tidal striping and mass segregation,
yielding an excess of very-low-mass stars in the outer parts of the cluster. 
Therefore it is not excluded that VLMS near $r_h$ may already have been stripped in this cluster.
In any case, it shows that {\it no} robust conclusion about the IMF of GCs can be inferred
at present from the observations of $NGC6397$, and probably not from any cluster located at
low Galactic latitude.

The core relaxation times of the present clusters are $\sim 10^8-10^9$ yr,
at least one order of magnitude less that the ages of the clusters, so that
we expect the stars near the central core of the clusters to be close to thermal equilibrium.
Therefore low-mass stars in the core should be substantially depleted as a result of mass segregation.
The half-mass radius is in general about 50 times larger than the core radius, so that we expect mass segregation to be mild in this region.
The effect of mass segregation in GCs has been demonstrated by various investigations
of the {\it core} ($r_c\approx 0.01\,r_h$)
of different clusters (DeMarchi \& Paresce, 1996; Paresce et al., 1995a;
King et al., 1995).
The severe deficiency of faint stars in the center of the clusters translates into a strongly decreasing MF
below $\sim 0.6\msol$. This
reflects the effect of mass segregation in the central core,
due to two-body relaxation, as expected from nearly thermodynamic equilibrium,
providing the time for core collapse is much larger than the half-mass
relaxation time, which is the case in general.
This result is confirmed by the analysis of King, Sosin \& Cool (1995) who examined
the relation between the {\it local} MF of the cluster,
as determined presently,
and the {\it global} MF, using a
dynamical model for the evolution of the cluster.
As shown by these authors, whereas the MF determined in the central core is
severely affected by mass segregation,
the MF determined
near the half-mass radius is quite similar to the global MF.


\subsection{Predictions in the NICMOS filters}

\begin{figure*}
\mbox{\epsfxsize=85mm
\epsfbox{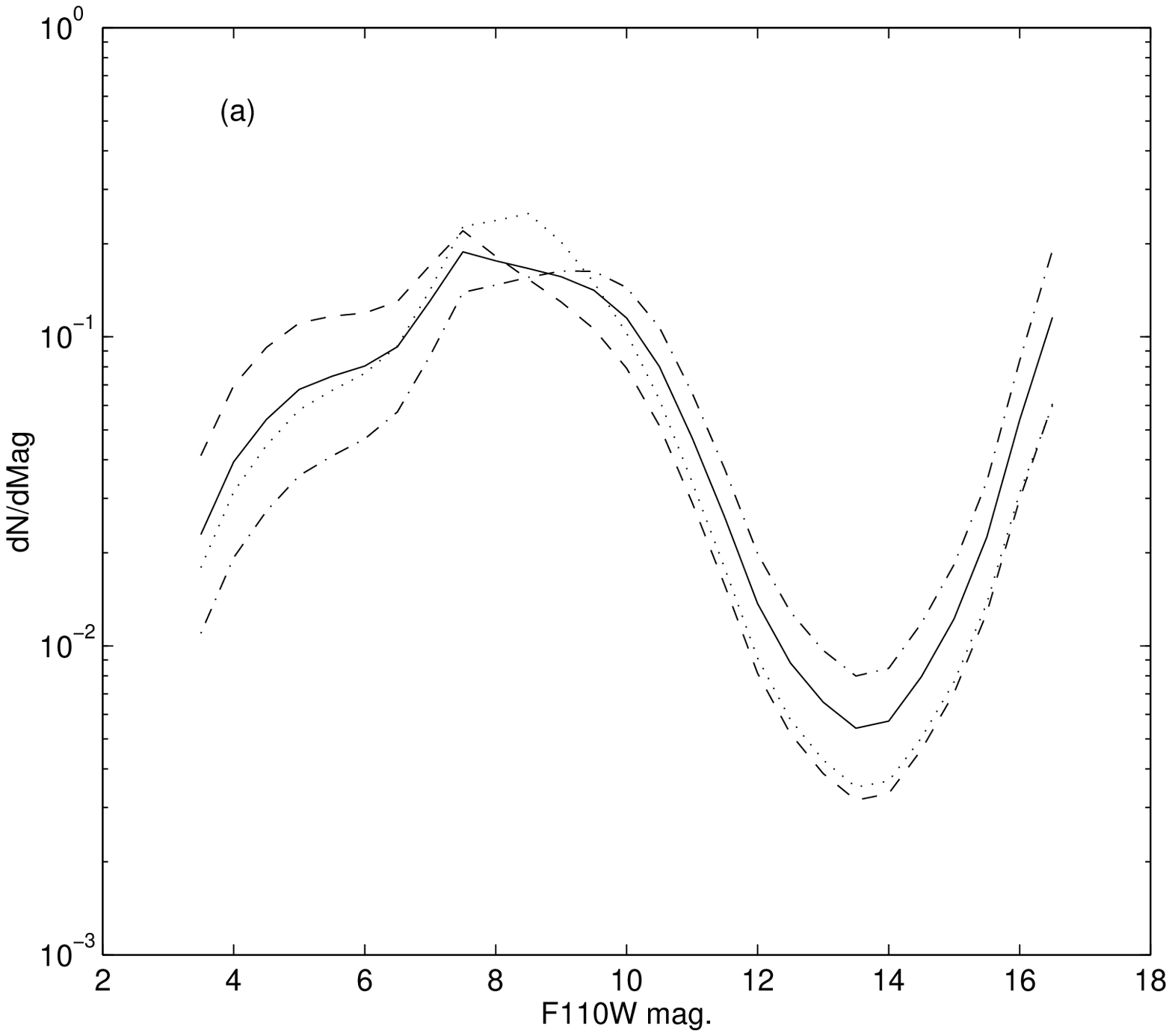}}
\mbox{\epsfxsize=85mm
\epsfbox{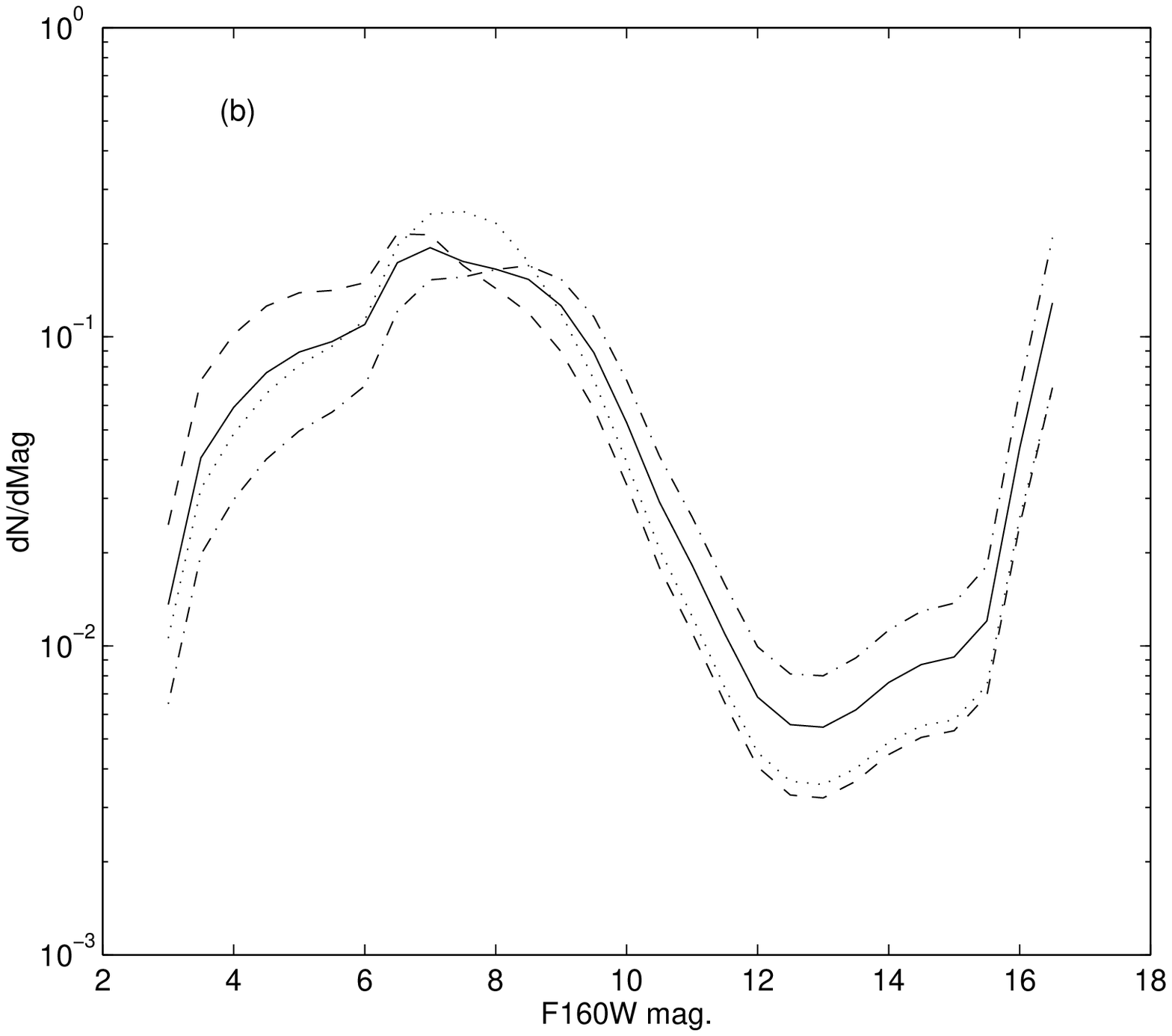}}
\mbox{\epsfxsize=85mm
\epsfbox{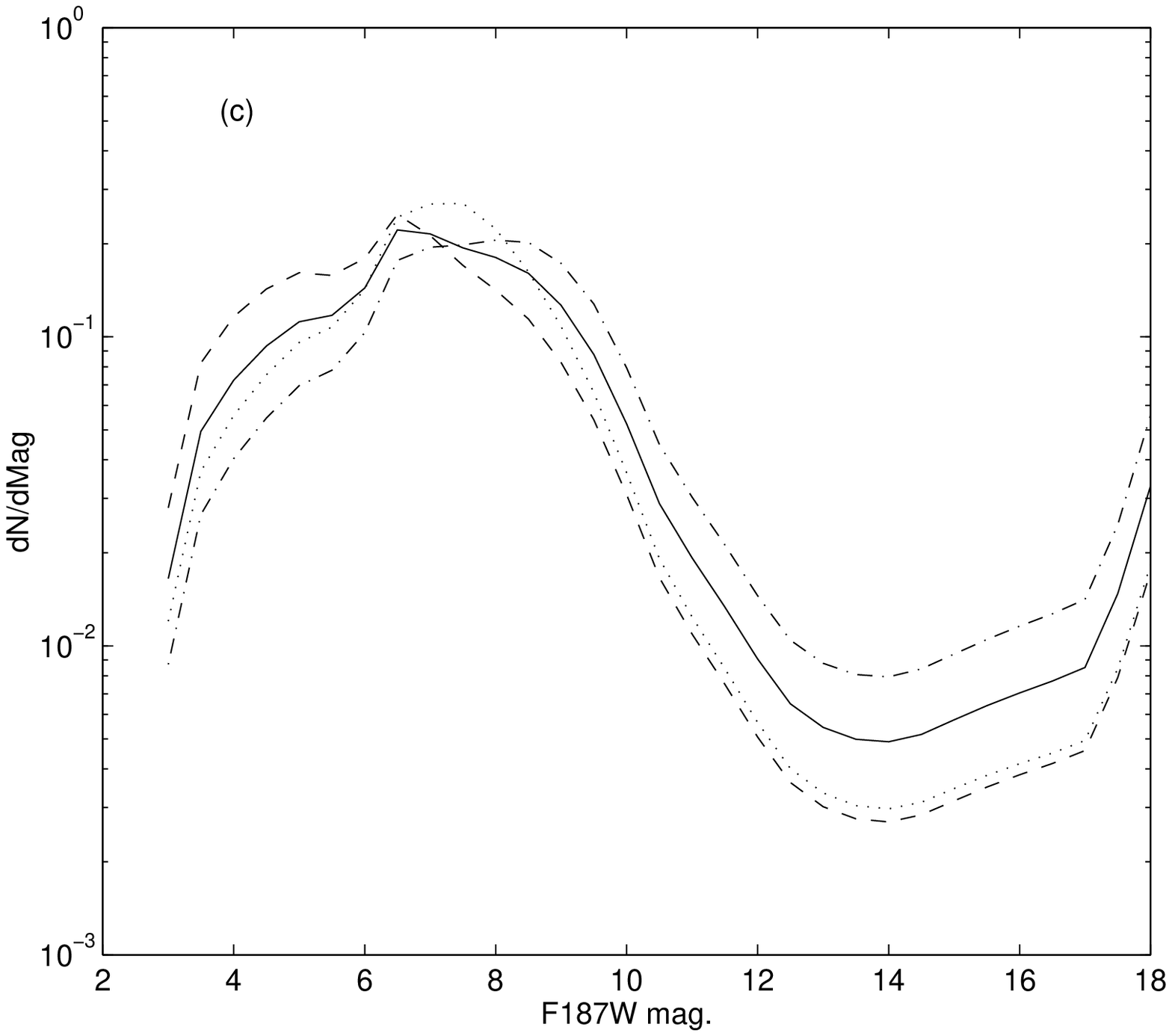}}
\hspace{0.4cm}
\mbox{\epsfxsize=85mm
\epsfbox{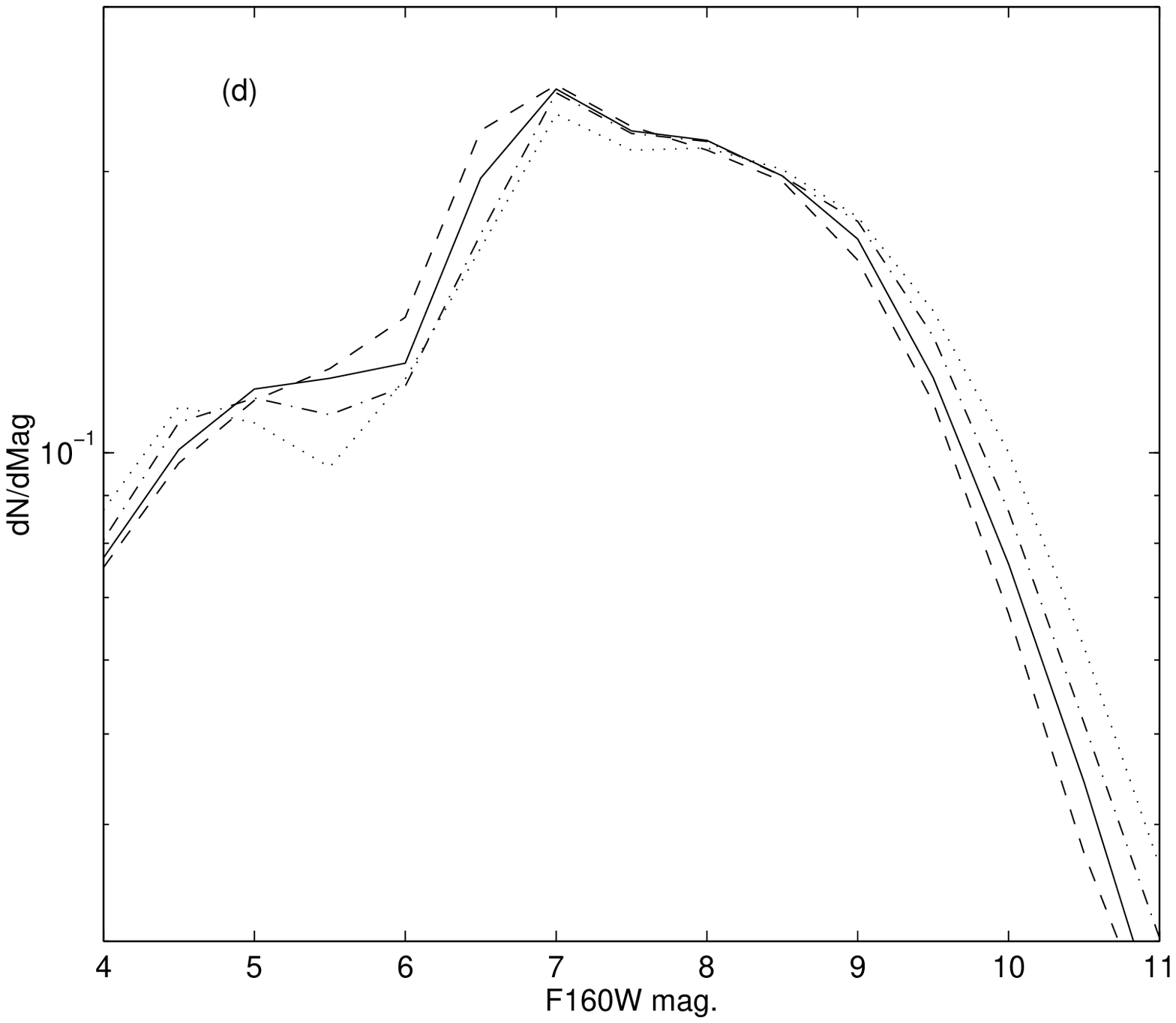}}
\caption[]{
Predicted LFs for clusters with $\mh=-2$ for different mass functions :
dotted line : $\alpha =0$ below 0.2 $\msol$; dashed line : $\alpha =0.5$; solid line : $\alpha = 1.0$; dot-dash line : $\alpha =1.5$
in the NICMOS filters F160W (Fig. 4a), F187W (Fig. 4b) and F110W (Fig. 4c).
The effect of metallicity is illustrated on Fig. 4d for $\alpha=1.5$ :
$\mh=-0.5$ (dotted line), $\mh=-1.0$ (dot-dash line), $\mh=-1.5$ (solid line)
and $\mh=-2.0$ (dashed line). 
}
\end{figure*}

The observations of GCs in near infrared colors will soon be possible with
the HST NICMOS camera, in the $F110W$ ($\sim J$), $F160W$ ($\sim H$) and $F187W$
filters, where $J$ and $H$ are defined in the CIT system.
Mass-magnitude relationships and CMDs have been derived by BCAH in these filters.
In Figures 4a-c, we show the predicted {\it luminosity functions} in these filters for $\mh=-2$,
as obtained with three different {\it rising} MFs ($\alpha >0$) and a flat MF
($\alpha=0$) beyond 0.2 $\msol$.
The LFs are normalized to $N_{\rm tot}=\int_{M_{\rm min}}^{M_{\rm max}} (dN/dm)\,dm=1$, where
$N_{\rm tot}$ is the total number of stars between 0.8 $\msol$, about the turn-off mass, and the hydrogen burning limit, which corresponds to magnitudes $M_{\rm min}$ and $M_{\rm max}$, respectively, for the appropriate metallicity, as given in BCAH.
The slope of the MF, at least for $\alpha > 0.5$, could be determined by future observations from differential
comparisons of the star counts on each side of the peak of the LF, {\it providing
accurate completeness corrections}. Steeper MFs yield more (resp. less) VLMS below
(resp. above) $\sim 0.2\,\msol$, which corresponds to the bright peak in the LFs. A flat MF yield a larger number of stars and
a small plateau between $\sim 0.1-0.2\,\msol$, followed by an abrupt drop.
The effect of metallicity on the LF is illustrated in Figure 4d for $\alpha = 1.5$. Although the qualitative behaviour of the LF remains
the same, the effect of metallicity is a global shift of the LF toward slightly brighter magnitudes with decreasing metallicity, a direct consequence of the metallicity dependence of the mass-magnitude relationship (see BCAH, Chabrier \& Baraffe, 1997a). The effect,
however, remains small and shows the weak dependence of the LF upon metallicity, at least for metal-depleted objects.

An other interesting feature about these LFs is the expected {\it brown dwarf} signature, as illustrated by the peak at faint magnitudes in Figs. 4a-c. The gap in the LF stems from the
large luminosity drop in the mass-luminosity relationship in the stellar to
substellar transition domain, which corresponds to several magnitudes over a few hundreths of a solar mass, whereas the sharp increase reflects the flatter slope of this relation in the brown dwarf regime.
We stopped our calculations at $m=0.055\,\msol$, which corresponds to the
coolest non-grey model atmospheres presently available. This is enough to illustrate the unmissing signature of even a moderate BD population in GCs,
as obtained with the mentioned MFs.
Interestingly enough, we note that the gap in the LF varies significantly with
the filters, which reflects the strong variation of the BD energy distribution
in the near infrared frequency domain. Even in the most favorable case,
i.e. $F110$, the genuine brown dwarf domain lies $\sim 4$ mag below the bottom
of the main sequence. This sets the scale for BD search in GCs.

\subsection{Brown dwarf mass fraction}

The mass under the form of brown dwarfs in GCs
is determined by :

\begin{eqnarray}
M_{\rm BD}=A \int_{m_{\rm low}}^{m_{\rm up}} m {dN\over dm} dm= {A\over 2-\alpha} (m_{\rm up}^{2-\alpha}
- m_{\rm low}^{2-\alpha})
\end{eqnarray}
where A is the normalization of the MF $dN/dm=A m^{-\alpha}$. We have normalized all the
MFs at $\sim 0.4-0.5$ $\msol$, as mentioned previously (see \S2).
The value $m_{\rm up}=0.083\,\msol$ is the hydrogen burning minimum mass for $\mh=-2$ and about it for metal-poor stars in general
(Chabrier \& Baraffe, 1997a) whereas $m_{\rm low}$ is the minimum mass of the brown dwarf IMF. Since this quantity is
presently unknown, the {\it maximum} brown dwarf contribution corresponds to $m_{\rm low}=0$, since the integral is convergent ($\alpha < 2$).
A more realistic value $m_{\rm low}=0.01\, \msol$, which corresponds to the deuterium burning
minimum mass (Saumon et al., 1996) and to the minimum Jeans mass at zero-metallicity (Silk, 1977) does not change the results significantly, given the small contribution of the lowest masses for weakly rising MFs.

Even the steepest MF $\alpha =1.5$ yields a brown dwarf contribution $\sim 30\%$ of the stellar mass fraction below
0.8 $\msol$, about the cluster turn-off mass.
A flat MF or a larger minimum mass $m_{\rm low}$ yield essentially no
mass ($< 1\%$) in the brown dwarf domain. These calculations show that,
even though there might be a substantial {\it number} of BDs in GCs (rising MF), their {\it mass} fraction is fairly small. This confirms that there is
essentially no dark matter in GCs.

\section{Halo mass function}

If the PDMF slope is not strongly correlated with the metallicity,
we expect the IMF for GCs to be rather similar to that of the
disk and the halo.
This issue is addressed below. 

As shown in \S 4.3 of BCAH, there is no apparent difference in the sequences of
globular star clusters and halo field stars\footnote{The term {\it halo} is
used for different populations in the literature and leads to some confusion. In the
present paper, we denote {\it stellar halo} or {\it spheroid} the stellar
population determined by the Hubble ($1/r^3$) or the DeVaucouleurs
($e^{-r^{1/4}}$) density profile, whereas we call {\it dark halo} the outer
$1/r^2$ density distribution. We will use the generic term {\it halo} to denote the
overall population belonging to the afore-mentioned spheroid {\it and}
dark halo, in opposition to the {\it disk} population.}.
This corroborates the traditional view that globular clusters and halo field stars are members
of the same stellar population (see e.g. Fall \& Rees, 1985).

The direct determination of the halo LF is a tremendously difficult task. A 0.09
$\msol$ star
with $\mh=-0.5$, about the metallicity of the old (thick) disk population
(see e.g. Leggett, 1992), has a magnitude $\mv\sim 17$ (Baraffe et al., 1995). Since the old disk scale height is $\sim$1-1.5 kpc
(Kuijken \& Gilmore, 1991; Bienaym\'e, Robin \& Cr\'ez\'e, 1987), we expect the disk population to contribute
significantly to star counts up to at least $V\sim 27$, $I\sim 23$.
This clearly illustrates the
difficulty of an accurate determination of the spheroid or halo LF from ground-based photometric measurements.
This difficulty is in principle circumvented with the Hubble Space Telescope, which reaches
$V\sim 30$ in the Hubble Deep Field (HDF) with 0.1 arcsec resolution, but the small field of the HDF
($\sim 5$ arcmin$^2$) yields too small statistics to derive a reliable halo LF.

This is why we must rely on the nearby halo LF, based on {\it geometric} parallax determinations
of
high-velocity faint stars in the solar neighborhood, as obtained by
Dahn et al. (1995). The high tangential velocity of the stars in the Dahn et al. LF 
($v_{\rm tan}\ge 220$ km.s$^{-1}$) is a strong indication of a halo population.
As shown in BCAH (Fig. 8), the sequence of high-velocity subdwarfs of Monet et al. (1992), part of the Dahn et al. (1995) sample, is fully consistent
with the GC sequences with an {\it average} metallicity $\mh\sim -1.5$
and a dispersion $\Delta \mh \approx \pm 0.5$ (BCAH). However, as shown in the same afore-mentioned figure of BCAH, a substantial fraction of the subdwarfs observed by Dahn et al. (1995), which all have halo-like kinematic properties, exhibit colors characteristic of metal-rich
($\mh \ga -0.5$) objects. The very nature of these stars remains a puzzle. Given the extremely low probability for a large number of disk objects to have such large velocity dispersions, these objects are likely to be halo stars formed in the Galactic central region. A consistent, bin by bin analysis of these observations requires precise completeness corrections, which are not available, so we can only derive a global analysis.
The luminosity of LMS increases with
decreasing metallicity, for a given mass (BCAH). Conversely, for a
{\it given magnitude}, in the range of interest, the mass of a LMS with
$\mh=-0.5$ is $\sim$0.1 $\msol$ larger than with $\mh=-1.5$. Therefore,
assuming a metallicity $\mh =-1.5$ for {\it all} the stars entering the
Dahn et al. LF will slightly overestimate the low-mass end of the MF and
thus provides an upper limit for the slope and the normalization of the
halo MF at the bottom of the MS.

The MF determined from the Dahn et al. LF for this metallicity is shown in Fig. 5 (solid line).
Also shown is the MF derived from the {\it photometric}
spheroid LF determined by Richer \& Fahlman (1992, RF) (dashed line). As shown in Fig. 5 (see also M\'era, Chabrier \& Schaeffer,
1996), both MFs are fairly consistent for $m\wig > 0.15\,\msol$ but
differ significantly below this limit.
The RF's MF rises abruptedly
whereas the Dahn et al. MF drops below $m\sim 0.10\, \msol$.
Such a qualitative difference is surprising, for both LFs are supposed
to probe the same stellar population, essentially the spheroid one.
As mentioned above (see also Reid et al., 1996), the last bin in the RF LF is likely to be contaminated by
{\it old disk} stars and can not be used to derive a halo MF.
The difference between the two MFs shown in Figure 5 becomes significant below $m\sim 0.15\,\msol$,
i.e. $\mv \wig >12$, $M_I\wig > 10$, $\mbol \wig >11$ for $\mh=-1.5$ (BCAH; Chabrier \& Baraffe, 1997a).
At 15 kpc this corresponds to $V\sim 28$, $I\sim 26$.
At faint magnitudes,
e.g. $V\sim 25$, an old-disk 0.4 $\msol$ star ($\mh \ge -1.0$; $\mv \sim 10$) at 10 kpc can be
misidentified as a metal-poor $\sim 0.1$ $\msol$ halo star ($\mh \le -1.5$;
$\mv \sim 15$) at 1 kpc with similar $V-I$. This will yield a severely overestimated halo number
density near the bottom of the MS. Note also that there are
only 3 stars in the two last bins of RF's LF.

On the other hand, the last 3 bins in the Dahn et al. LF rely on a fairly small number of stars
(7, 5 and 1, respectively), and have a low statistical significance.
At last we want to emphasize the bias introduced by bining the data at faint magnitudes.
The number of stars per unit magnitude
is about 5 times smaller below $\sim 0.1\,\msol$ than above this value (see e.g. BCAH), because of the abrupt drop in the
mass-luminosity relationship. A good statistics thus implies a large magnitude interval but this
in turn yields a large indetermination in the exact magnitude of the
stars belonging to the interval.
The faint end of the Dahn et al. LF is likely to be hampered
by these limitations, due mainly to incompleteness and to
the scarcity of objects
in a large magnitude range at the very bottom of the main sequence.
Note that the same observational bias could apply to the uncertain determination of the faint
end of $NGC6397$ LF, as discussed in the previous section.

\begin{figure}
\epsfxsize=88mm
\epsfbox{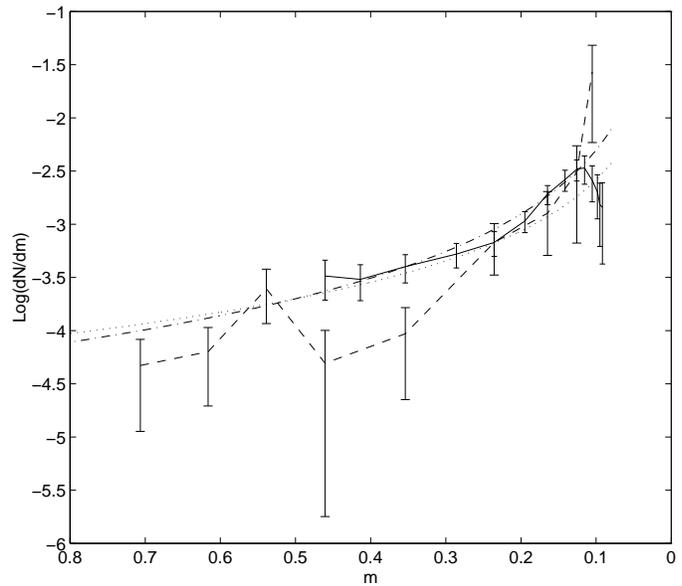}
\caption[]{
Mass function for halo field stars. Solid line : MF determined from the nearby LF of high-velocity
subdwarf (Dahn et al., 1995); dashed line : MF determined from the photometric LF of Richer \&
Fahlman (1992), both for $\mh =-1.5$. Dot and dash-dot lines : power-law MFs with $\alpha=1.6$ and $\alpha=2.0$, respectively.
}
\end{figure}

As shown in Fig. 5, the Dahn et al. MF is consistent with a rising MF
$dN/dm \propto m^{-\alpha}$ with $\alpha \sim 2$ down to $0.12 \,\msol$
at the 1-$\sigma$ (99\%) confidence level, and  with $\alpha \sim 1.6$ down to the very last bin, i.e.
0.09 $\msol$ at the 2-$\sigma$ (90\%) C.L., for $\mh=-1.5$. This C.L. increases
if we use a slightly lower metallicity. A $\chi^2$-test applied to the MF determined in Figure 5
yields a most-likely slope $\alpha \sim 1.64$, assuming a power-law MF,
although we cannot exclude a non power-law behaviour at the very end of the MF.
Interestingly enough, similar conclusions have been reached recently by Richer \& Fahlman (1996) who reanalyzed their photometric LF and derived a MF with a
slope $\alpha \sim 1.7$.
A more precise determination of the shape of the MF near the hydrogen-burning
limit, which bears essential consequences for star formation theory, requires
larger statistics in this region. However the present determination is sufficient
to determine with reasonable accuracy the halo stellar and substellar mass
fractions, as examined in the next section.

\section{Halo stellar and substellar mass fractions}
 
The spheroid extends between $\sim 10$ and 20 kpc from the Galactic center
(see e.g. Reid et al., 1996 and below),
i.e. $V\sim 25- 30$,
$I\sim 23-28$
for the bottom of the MS of metal-poor stars (BCAH).
This is out of reach with ground-based observations
but can be probed partially with the HDF, which reaches $V\sim 30$, $I\sim 28$
(Flynn, Gould \& Bahcall, 1996).
As shown in the previous section, a power-law MF $dN/dm\propto m^{-\alpha}$ with a slope $\alpha \sim 1.6-2.0$ is fairly consistent with all halo
MFs (excluding the last bin of RF) down to at least $\sim 0.12\, \msol$,
at the 99\% C.L.,
and down to the limit of the observations at the 90\% C.L.
This yields the following halo MF :

\begin{eqnarray}
\mu_h(m)\approx 4.0 \pm 2 \,\times 10^{-3} ({m\over 0.1\,\msol})^{-(1.7 \pm 0.2)}\,\, \msol^{-1}pc^{-3}
\end{eqnarray}

Integration of this MF gives the halo stellar ($m\le 0.8 \,\msol$ for $t\ge 10$ Gyr) density in the solar neighborhood:

\begin{eqnarray}
\rho_{h_{\star}}\approx 1.0 \times 10^{-4}\,\msol.pc^{-3}
\end{eqnarray}

Comparison of this value with the {\it disk} low-mass star density (M\'era,
Chabrier \& Baraffe, 1996), and assuming that the MF (2) is valid for masses up
to $m=0.8\,\msol$ ($\mv \sim 4)$ gives the
mass-normalization of the halo/disk total stellar population (including the contribution of more massive stars for the disk population) down to the bottom of the MS, $\rho_{h_\star}/\rho_{d_\star} \sim 1/400$.

\subsection{Star count analysis}

Although the small field of the Hubble Deep Field prevents the determination
of an accurate halo LF, as mentioned previously, it provides a stringent constraint on the halo mass function and mass model.

The number of stars per arcmin$^2$ between apparent magnitudes e.g. $I_{\rm min}$ and $I_{\rm max}$ toward Galactic coordinates ($b,l)$  reads :

\begin{eqnarray}
{dN\over d\Omega} & = & {1\over 3600}({\pi\over 180})^2 \nonumber \\
& \times & \int_{M_{\rm min}}^{M_{\rm max}}dM_I\,
\int_{d_{\rm min}}^{d_{\rm max}} dr\,  r^2 \rho\bigl (r(l,b)\bigr )
{dN\over dm} {dm\over dM_I}
\end{eqnarray}
\noindent where $d_{\rm max}/10 \,pc=10^{{I_{\rm max}-M_I\over 5}}$ (resp. $d_{\rm min}$) denotes the
maximum (resp. minimum) distance to which a star of magnitude $M_I$ will be visible, $M_{\rm min}$-$M_{\rm max}$ the magnitudes corresponding to the mass range of interest ($\le 0.8\,\msol$) and $\rho(r)$ is the mass density profile characteristic of the Galactic region considered. The mass-magnitude relation $dm/dM_I$ is taken from BCAH.
Table II gives the predicted star counts (between 0.8 $\msol$ and the hydrogen
burning limit) at faint magnitudes in the Hubble Deep field,
$l=126^o$, $b=55^o$, for the observed field of view, 0.00157 sq. degrees.
The Galactic model consists of a {\it thick disk} with an exponential $\propto e^{-|z|/z_0}$ or a $\propto \mbox{sech}^2{z\over z_0}$
profile with different scale heigths, $z_0=1000$ and 1500 pc, a {\it spheroid} with a DeVaucouleurs $\propto e^{-10(R/R_\odot)^{1/4}}$ profile and a flattened ($q=0.6$) {\it dark halo} with an isothermal distribution $\rho(r)\propto 1/r^2$. The mass functions are the ones determined in M\'era et al. (1996b) for the disk and in Eq. (2) for the halo. The average metallicity is $\mh = -0.5 $ for the disk and $\mh = -1.5 $ for the halo, respectively.
As shown in Table 2, the results obtained with spherically symetric distributions are in severe
conflict with the observations, whereas the observed star counts are consistent
with an oblated $q=c/a\sim 0.6$ spheroid plus a sech$^2$ thick disk
with a scale height $z_0\wig < 1$ kpc. The observations of the HDF exclude completely the presence of
a significant {\it dark halo} stellar population. Therefore, almost {\it all}
the stars in the Dahn et al. sample belong to the {\it spheroid}.

\bigskip
\begin{table*}
\caption[]{Different contributions to the star counts in the 5.6 Arcmin$^2$ Hubble Deep Field $(b,l)=(55^o,126^o)$ (Reid et al., 1996) with
the mass function determined in M\'era et al. (1996b) for the disk and in Eq. (2) for the halo. 
We considered different density-profiles (see text).
The repartition between disk and
halo in the observations is just indicative and has been obtained from color criteria
(see e.g. Reid et al., 1996 Fig. 13).}
\begin{center}
\begin{tabular}{lll}
\hline\noalign{\smallskip}
                   & $I=20-26$ & $I=24-26$   \\
\noalign{\smallskip}
\hline\noalign{\smallskip}
 {\bf old disk}    &     &        \\
 $e^{-|z|/z_0}$; $z_0=1.0$ kpc &   9  & 2.6     \\
 $e^{-|z|/z_0}$; $z_0=1.5$ kpc &  23  & 8.7    \\
 sech$^2({z\over z_0})$; $z=1.0$ kpc &  3.9   & 0.7     \\
 sech$^2({z\over z_0})$; $z=1.5$ kpc &  12.2   &  3.2   \\ \hline
\noalign{\smallskip}
 {\bf Observations}    &  4-6   &  1-2      \\ \hline
\noalign{\smallskip}
 {\bf spheroid}    &     &       \\
 $q=1$             &  12.3   &  5.4     \\
 $q=0.6$           &  5.8   &  1.8     \\ \hline
 {\bf dark halo}   &     &       \\
 $q=0.6$           &  244   &  176    \\ \hline
 {\bf Observations}   &  5-7   &  3-4      \\
\noalign{\smallskip}
\hline
\end{tabular}
\end{center}
\end{table*}


An event more stringent comparison is given in Table III which compares the predicted star counts with the HDF observations
in two different passbands, namely the V-band (Mendez et al., 1996) and the I-band (Reid et al., 1996). The Galactic model consists of the afore-mentioned {\it thick disk} characterized by a density profile $\rho(r)=
0.05 \,\rho_\odot\,e^{-(R-R_\odot)/l_{od}}\, \mbox{sech}^2{|z|\over
z_0}$ with a local density $\rho_\odot\approx 0.1\,\msvol$, a solar
galactocentric distance $R_\odot=8.5$ kpc, a scale length $l_{od}=
2.8$ kpc and a scale height $z_0= 1$ kpc, and a flattened ($q=0.6$) DeVaucouleurs {\it spheroid}. The spheroid
and disk populations are determined from their location in the CMD (see e.g. Reid et al., 1996 Fig. 13; Baraffe et al., 1995). We note
the excellent agreement (within $\sim 1\sigma$) between theory and observations in the two
passbands, for {\it every} color and magnitude interval.
The spheroid VLMS
population extends up to $I\sim 27,\, V\sim 29$ and vanishes beyond this
limit. Most of the spheroid stars in the range $25<I<27$, $1.5 \wig <
V-I \wig < 2.5$ have masses between 0.1 and 0.2 $\msol$, as determined
from the present stellar models and mass-color relations for different
metallicities (BCAH). The scarcity of stars at fainter magnitudes thus reflects
{\it the bottom of the spheroid MS}. The absolute magnitude of a 0.1
$\msol$ star with $\mh\sim -1.5,-2.0$ is $M_V\sim 13.5,\,M_I\sim 11.3$
(BCAH) so that these stars are located at $\sim 10- 15$ kpc from
the Sun. The present analysis provides a relatively precise determination of the size of the Galactic spheroid. It also shows the consistency of the present disk and halo mass-functions and Galactic model with HDF faint magnitude observations
in different colors. It also confirms the negligible stellar population in the
dark halo, at most $\sim 1\%$ of the {\it spheroid} population, as inferred from Table II. The present analysis shows that {\it we have seen essentially all main sequence stars in the Galaxy}.

\bigskip
\begin{table*}
\caption[]{Star counts in the Hubble Deep
Field in the V-band (Mendez et al., 1996; 4.69 arcmin$^2$) and in the I-band (Reid et al., 1996; 5.6 arcmin$^2$) (tabulated CMDs). See text for the Galactic model. The average metallicity is $\mh = -0.5 $ for the disk and $\mh = -1.5 $ for the spheroid, respectively. The numbers in brackets for the spheroid correspond to
$\mh=-2$. 
At fainter magnitudes, the predicted star counts are consistent with zero in the color-range of interest.
}
\begin{center}
\begin{tabular}{llllllllllllll}
\hline
\noalign{\smallskip}
I\,\,/\,\,(V-I)        & 0.5-1.5 & 1.5-2.5 & 2.5-3.5 & & & & & & & V\,\,/\,\,(V-I)        & 0.5-1.5 & 1.5-2.5 & 2.5-3.5 \\ 
\hline\hline\noalign{\smallskip}
  18-20             &  & &  &  & &  & & & & 21-23 &  & &  \\
\,\, Thick disk     & 0.1 & 0.8 & 0.2  &  & &  & & & & & 0 & 1.2 & 0.1 \\
\,\, Spheroid       & 0.4\,(0.5) & 0.1\,(0) & 0\,(0)  & &  & &  & & & & 0.5\,(1) & 0.5\,(0.3) & 0\,(0)\\
\,\, Obs.           & 1 & 2 & 0  &  & &  & & & & &  2 & 0 & 0 \\
  20-22             &  & &  &  & &  & & & & 23-25 &  & &  \\
\,\, Thick disk     & 0 & 1.4 & 0.3  &  & &  & & & & & 0 & 0.9 & 0.6 \\
\,\, Spheroid       & 0.6\,(1) & 0.6\,(0.3) & 0\,(0) & &  & &  & & & &  0.3\,(0.8) & 1.5\,(1.1) & 0\,(0)  \\
\,\, Obs.           & 2 & 1 & 1  &  & &  & & & &  & 1 & 2 & 1 \\
  22-24             &  & &  &  & &  & & & & 25-27 &  & &  \\
\,\, Thick disk     & 0 & 0.6 & 0.9 &  & & &  & & & &  0 & 0.1 & 0.7  \\
\,\, Spheroid       & 0.3\,(0.8) & 1.7\,(1.3) & 0.1\,(0) & &  & &  & & &  & 0\,(0.2) & 1.6\,(1.4) & 0.1\,(0.1)  \\
\,\, Obs.           & 2 & 2 & 1  &  & &  & & & & & 1 & 1 & 0 \\
  24-26             &  & &  &  & &  &  & & &  & & &  \\
\,\, Thick disk     & 0 & 0 & 0.3  &  & &  &  &  & &  & & &  \\
\,\, Spheroid       & 0\,(0.2) & 1.6\,(1.5) & 0.3\,(0.2) & & & & & & & & & & \\
\,\, Obs.           & 0 & 2 & 1  &  & &  & & &  &  & &  &  \\
\noalign{\smallskip}
\hline\hline
\end{tabular}
\end{center}




\end{table*}

The spheroid stellar density (3) is thus
about 1\% of the dynamically-determined dark matter density and 0.1\% of the disk {\it total}
density, whereas the dark halo stellar density is about two orders of
magnitude smaller, as implied by the present star-count analysis. This yields a
contribution of the spheroid to the optical depth
towards the LMC $\tau_{\rm sph} \sim 5\times 10^{-9}$, the eventual dark-halo contribution
being negligible.

The extrapolation in the BD domain, as done in the previous section, yields a negligible amount
of substellar objects in the Galactic halo.
A minimum mass $m_{\rm low}=0.01\,\msol$ for the BDs and the extrapolation of the MF (2)
yields a BD number-density in the spheroid:

\begin{eqnarray}
n_{h_{BD}}\sim 2\pm 1\times 10^{-3}\, \mbox{pc}^{-3},
\end{eqnarray}
\noindent and a BD mass-density :

\begin{eqnarray}
\rho_{h_{BD}}\sim 0.5\pm 0.5\times 10^{-4}\,\msol \mbox{pc}^{-3},
\end{eqnarray}
\noindent i.e. $\wig < 1 \%$ 
of the dark
matter density.
This is consistent with a detailed analysis of the MACHO+EROS microlensing observations toward the
LMC (Alcock et al, 1996; Aubourg et al., 1993; M\'era, Chabrier \& Schaeffer, 1997). A substantial fraction of
halo {\it brown dwarfs} not only would imply a MF below the hydrogen burning
limit radically different (much steeper) from the one determined by Eq. (2), but would also
imply a
large rate of short-time ($\wig <20$ days) events ($\Gamma \propto 1/\sqrt m$),
whereas {\it no} event beyond 19 days has been observed by either the MACHO or the EROS experiments towards the LMC. It confirms that the last bin of the RF LF,
which would yield $\alpha \sim 5$ is not correct.
This clearly
excludes brown dwarfs as plausible candidates for the halo missing mass and for
the observed events towards the LMC.

Interestingly enough, we note that the MF (2) has a slope very similar to
the one determined previously for
the {\it disk} population down to the bottom of the MS, $1.5\wig < \alpha \wig < 2.5$ (M\'era, Chabrier \& Baraffe, 1996). The small number statistics at faint magnitudes prevents a more precise determination. Note that
such a {\it rising} MF is consistent with microlensing observations towards the central bulge
whereas a decreasing MF near the hydrogen burning limit would be in severe conflict ($> 5\sigma$) with
these observations (Han \& Gould, 1996; M\'era et al., 1997).
If confirmed by future observations, this strongly suggests that the
stellar MF is a rather universal property, almost independent on the metallicity, and reflects the
nature of a dominant physical process in star formation.
It would also show that a Salpeter-like MF extends to about three orders of
magnitude in mass, then justifying the extrapolation into the brown dwarf domain. Note that the metallicity might be determinent for the {\it minimum mass} of the MF.

\section{Conclusion}

In this paper, we have first transformed the LFs of several globular clusters observed with the HST in various passbands down to almost the bottom of the main sequence into {\it bolometric} LFs. {\it All} these LFs exhibit a similar behaviour with a peak around $\mbol \sim $9-10 and a drop beyond. This reflects the sharp drop in the {\it mass-luminosity relationship} below
$\sim 0.2 \,\msol$. We have converted these LFs into {\it mass functions}.
We emphasize that these mass functions rely on mass-luminosity
relationships obtained with stellar models whose accuracy has been
demonstrated unambigously (BCAH).
Since the luminosity functions for these clusters have been determined near the
half-mass radius, the present-day local mass functions are likely to reflect reasonnably well
the {\it initial} global mass functions. All the MFs are rising down to $\sim 0.1\,\msol$,
with the exception of the one based on the observations of $NGC6397$ by
Paresce et al. (1995) and possibly the one for $47Tuc$ which seem to flatten off below $\sim 0.2 \,\msol$. This stresses the need for further observations of these clusters to
determine whether the effect is real or whether it stems from various observational bias at faint magnitudes,
as suggested, for $NGC6397$, by comparison with the observations of the same cluster by Cool et al. (1996).
All these MFs are well described by a power-law
$dM/dm\propto m^{-\alpha}$, with essentially no dependence of $\alpha$ on metallicity or the Galactic location, with $\alpha$ varying between $\alpha \sim 0.5$ and $\alpha\sim 1.5$. The clusters thus seem to exhibit the same universal
{\it initial} mass function, shaped to some extent (within about one unit for the slope $\alpha$) by further dynamical evolution.
The effect of metallicity appears in the mass-magnitude relationship and
translates into a glogal shift of the LF, and thus of the location of the peak, which corresponds to $m\sim 0.2\,\msol$, to slightly brighter magnitudes for lower metallicity.
Given the completely different dynamical history of these clusters, this confirms that
the stellar population near the half-mass radius is weakly affected by either
external or internal dynamical effects. Such a conclusion would have to be reconsidered if the observations of Paresce et al. (1995) for $NGC6397$ are confirmed.
In that case, the Galactic latitude and the Galactocentric distance of GCs might
play a determinent role below $\sim 1$ kpc.
These weakly rising MFs suggest
that the globular clusters contain a substantial {\it number} of brown dwarfs
but a negligible amount of {\it mass}, and thus of dark matter, under the form of these objects.

The mass segregation effect is illustrated by the severe depletion of very low-mass stars
in the central core of these clusters, as reflected by the dropping MF below
$\sim 0.6\, \msol$, and by the
excess of these objects, as shown by the substantially steeper MF, in the outer
field of $NGC6397$.

We have derived predicting LFs in the NICMOS filters, corresponding to either
slightly rising or flat MFs for the different clusters, down to the brown dwarf domain. Comparison between these predictions and near-future observations should
determine accurately the MF of globular clusters all the way down to the hydrogen burning limit and the brown dwarf population.

We have applied these calculations to the determination of the {\it halo} mass function, from
the observed luminosity functions of high-velocity low-mass stars in the stellar neighborhood and from the direct photometric determination at faint magnitudes.
Both MFs agree fairly well down to $\sim 0.15\,\msol$. Such a consistency
adds support to the fact that they
probe the same stellar population.
Both type of observations lead to the same rising MF down to $\sim 0.15\,\msol$, but to two different
behaviours below this limit. However, both LFs at the faint end are hampered by
different observational bias and most likely are consistent with the same,
slowly rising MF all the way down to the hydrogen-burning limit with a slope
$\alpha \sim 1.7 \pm 0.2$, although a decreasing MF below $\sim 0.15\,\msol$ cannot be excluded with presently available observations.
The drop
in the LF reflects the
scarcity of objects over large magnitude intervals near the very bottom of the main sequence.

Comparison with star counts at faint magnitudes obtained with the Hubble Deep
Field are consistent with the afore-mentioned halo MF and with a $sech^2$ old disk with a scale height $z_0\wig < 1$ kpc, which contributes at least $\sim$ 50\% of the LF down to $I\sim 25$, $V\sim 27$. These comparisons
yield an independent determination of the size $\sim 10-15$ kpc and the oblateness $q\sim 0.6$ of the spheroid. These calculations show that the observed high-velocity subdwarfs belong to the Galactic
{\it spheroid}, which dominates the LF up to $V\sim 29$,
$I\sim 27$, and that there is essentially {\it no} main sequence star in the {\it dark halo}.
Comparison
between future deep field ($I>28$) observations and these calculations should determine
unambigously the dark halo stellar and substellar mass content.

These calculations yield a consistent determination of the spheroid MF
down to $\sim 0.1\,\msol$, and its normalization w.r.t. the disk one.
They also give an upper bound for the dark-halo normalization ($\wig < 1\%\,
\rho_{\rm sph}$). They yield the contribution of the spheroid and the dark-halo contribution to the dark matter density in the solar neighborhood, less
than $\sim 1\%$ and 0.01\%, respectively.
Extrapolating these mass functions into
the brown dwarf domain reveals the {\it negligible contribution of substellar objects to halo dark matter}, which is consistent with the lack of short time events in the microlensing observations towards the LMC.

The mass functions determined for globular star clusters and for halo field stars have the same
qualitative and semi-quantitative behaviour, which seems to confirm the common origin
of these stellar populations. More strikingly, these MFs are similar to
the one determined for the disk stellar population down to the bottom of the MS.
This general behaviour of the mass function for different populations down to
the hydrogen burning limit, in particular the weak correlation with the
metal content, seems to reflect a universal feature of all stellar systems
and thus is likely to stem from a fundamental mechanism in the star formation
process. Although the precise determination of the shape of the MF requires more
statistics, the present determination represents already a solid ground to infer
the halo stellar and substellar {\it mass content}, as done in the present study.
Note that the effect of metallicity might come into play in the determination of the
{\it minimum mass} for the formation process. Although this minimum mass seems to be close to $\sim 1\,\msol$
for dark halo objects, as inferred from the microlensing observations (Chabrier, Segretain \& M\'era, 1996; Adams \& Laughlin, 1996), which is consistent with the present calculations which exclude a significant low-mass star population in the dark halo,
it lies well below this limit for spheroid and disk objects (M\'era et al., 1997).

\medskip
The bolometric LF's displayed in Fig. 1 are available at CDS via anonymous ftp: 
\par
\hskip 1cm ftp cdsarc.u-strsbg.fr (130.79.128.5) \par
or via http://cdsweb.u-strsbg.fr/Abstract.html
\bigskip
\bigskip

\bigskip

\begin{acknowledgements} 
The authors are very grateful to A. Cool, R. Elson, G. DeMarchi, G. Gilmore and C. Dahn
fo kindly providing their data under electronic form and to M. Cr\'ez\'e for helpful conversations.
\end{acknowledgements}

\vfill
\eject

\end{document}